\documentclass[aps,reprint,superscriptaddress,prl]{revtex4-1}
\usepackage[utf8]{inputenc}
\usepackage{amsmath}
\usepackage{amssymb}
\usepackage{graphicx}
\usepackage{graphics}
\usepackage{color}
\usepackage[bookmarks=false]{hyperref}
\usepackage[usenames,dvipsnames]{xcolor}
\usepackage{subfigure}

\def\be{\begin{equation}}       \def\ee{\end{equation}}
\def\bea{\begin{eqnarray}}      \def\eea{\end{eqnarray}}
\def\ba{\begin{array} }
\def\ea{\end{array} }
\def\bnum{\begin{enumerate} }
\def\enum{\end{enumerate}}

\def\=>{\Rightarrow}
\def\>{\rightarrow}

\def\eye2{Fathbb{I}}

\def\d0{\Delta_{0}}

\def\Co122{ BaFe$_{2-x}$Co$_x$As$_2$}

\begin{document}


\title{Disorder Induced Suppression of CDW Long Range Order: STM Study of Pd-intercalated ErTe$_3$}


\author{Alan Fang}
\affiliation{Stanford Institute for Materials and Energy Sciences,\\
SLAC National Accelerator Laboratory, 2575 Sand Hill Road, Menlo Park, CA 94025}
\affiliation{Geballe Laboratory for Advanced Materials, Stanford University, Stanford, CA 94305}
\affiliation{Department of Applied Physics, Stanford University, Stanford, CA 94305}

\author{Joshua A. W. Straquadine}
\affiliation{Geballe Laboratory for Advanced Materials, Stanford University, Stanford, CA 94305}
\affiliation{Department of Applied Physics, Stanford University, Stanford, CA 94305}

\author{Ian R. Fisher}
\affiliation{Stanford Institute for Materials and Energy Sciences,\\
SLAC National Accelerator Laboratory, 2575 Sand Hill Road, Menlo Park, CA 94025}
\affiliation{Geballe Laboratory for Advanced Materials, Stanford University, Stanford, CA 94305}
\affiliation{Department of Applied Physics, Stanford University, Stanford, CA 94305}

\author{Steven A. Kivelson}
\affiliation{Stanford Institute for Materials and Energy Sciences,\\
SLAC National Accelerator Laboratory, 2575 Sand Hill Road, Menlo Park, CA 94025}
\affiliation{Geballe Laboratory for Advanced Materials, Stanford University, Stanford, CA 94305}
\affiliation{Department of Physics, Stanford University, Stanford, CA 94305}

\author{Aharon Kapitulnik}
\affiliation{Stanford Institute for Materials and Energy Sciences,\\
SLAC National Accelerator Laboratory, 2575 Sand Hill Road, Menlo Park, CA 94025}
\affiliation{Geballe Laboratory for Advanced Materials, Stanford University, Stanford, CA 94305}
\affiliation{Department of Applied Physics, Stanford University, Stanford, CA 94305}
\affiliation{Department of Physics, Stanford University, Stanford, CA 94305}


\date{\today}

\begin{abstract}
Pd-intercalated ErTe$_3$ is studied as a model system for the interplay between a bidirectional two component charge density wave (CDW) state and disorder. Using scanning tunneling microscopy (STM), we show that introducing Pd-intercalants (i.e. disorder) disrupts the long-range order of both CDW states via the creation of dislocations, which appear associated with each CDW separately. While for weak disorder both CDW states continue to coexist throughout the sample, with no ``domains'' of one CDW direction or another, increasing Pd concentration has a stronger effect on the secondary CDW state, manifested in higher density of dislocations. Vestiges of the two distinct CDW phases persist to intercalation levels much above where signatures of the original phase transition are totally suppressed. This study therefore presents a first look into the disruption of multiple 2D strong-coupling CDW states by the proliferation of dislocations.
\end{abstract}


\maketitle

\noindent {\bf Introduction} - 
Charge ordered states are a key feature in strongly correlated materials \cite{FradRMP2015,Keimer2015,KivRMP2003}.  For example, the cuprate high temperature superconductors exhibit various signatures of charge-ordering and fluctuating order, raising the question of their impact on the occurrence and nature of superconductivity.  While notionally ``charge order'' refers to states that spontaneously break the spatial symmetries of the host crystal, the presence of unavoidable quenched disorder, even in the cleanest materials, frequently precludes true long-range order.  This, and the fact that the magnitude of the resulting ion displacements are relatively small, delayed the identification of such order in the cuprates with traditional methods, such as X-ray scattering, even though evidence of local order was deduced earlier from Scanning Tunneling Microscopy (STM) and Spectroscopy (STS) studies \cite{Howald2003,Hoffman2002,McElroy2005}. Moreover, since quenched disorder always precludes any long-range-ordered incommensurate charge density wave (CDW) state  \cite{ImryMa1975}, it proved difficult to infer the exact form of symmetry breaking that would take place in the ``ideal,'' zero disorder limit in these systems \cite{Robertson2006,DelMaestro2006}, a question that may have far reaching implications for the nature of the electronic state in many other correlated electron systems.  

More generally, because charge order is so sensitive to quenched randomness, it is extremely important to complement spatially averaged information obtained from transport or diffraction measurements, with information from local probes.  Thus, to shed light on this issue, we turn here to a model system which mimics certain aspects of the main features of the charge ordering phenomena of the cuprates, where a fairly solid understanding of the theory of the pure system exists (see \cite{Nie2014} and supplementary Information of \cite{Jang2016}), and disorder can be introduced in a controlled fashion and its effects studied using both global (scattering and transport) and local (STM) probes. 

It was recently suggested that Pd-intercalated RTe$_3$ (R = rare earth element) might be such a model system \cite{He2016,Lou2016,Straquadine2019}. Here we explicitly study the case R = Er. The pristine (un-intercalated) compound undergoes two successive mutually perpendicular unidirectional incommensurate CDW phase transitions, with critical temperature $T_{CDW1} = 270$ K, and $T_{CDW2} = 165$ K respectively \cite{Ru2008}. (At low temperatures, this phase with two co-existing, but different, CDWs is called bidirectional.)  In particular, Straquadine {\it et al.} demonstrated that signatures of the two phase transitions in pristine ErTe$_3$ are rapidly smeared and suppressed by Pd intercalation, consistent with a scenario in which the dominant effect arises from disorder induced by the intercalant atoms \cite{Straquadine2019}. 

In this paper we present a low-temperature scanning tunneling microscopy study of the effects of disorder on the two orthogonal components of the incommensurate CDW order in Pd-intercalated ErTe$_3$. This allows us to obtain clear insight into the nature of the interplay between the two components of the order, the relation between the fundamental density wave order and composite orientational (``vestigial nematic'' \cite{Nie2014}) order, and the role of topological defects (dislocations) in all these phenomena. Our principal results are: 
{\it i}) Both of the two perpendicular CDWs resolved by the STM are consistent with bulk measurements \cite{Moore2010,Straquadine2019}, including the fact that the wavelengths in the two directions are slightly but significantly different from each other.
{\it ii})  Both CDW components coexist throughout the sample, whether in the absence or presence of disorder, consistent with the intrinsically bidirectional character of the low temperature CDW order. However, the CDW associated with the larger gap and $q_{CDW1} = 0.70c^*$ is dominant.
{\it iii})  Introducing Pd-intercalants disrupts the long-range order of both components of the CDW via the creation of both phase disorder and dislocations (primarily dislocation pairs), which appear associated with each component separately. This has a stronger disruptive effect on the secondary CDW, which has a greater number of dislocations than the dominant component. 
{\it iv}) In the presence of disorder, nematic order is preserved over much longer length scales than the CDW order itself, consistent with there being nematic long range order \cite{Nie2014,Jang2016}.  For example, at weak disorder (our lowest intercalation sample of 0.3\%), there are typically no dislocations in the ``stronger'' component of the CDW order (within the scan range), which we suggest indicates the first observation of a ``nematic Bragg Glass'' phase \cite{Feigelman1989,Nattermann1990,Bouchaud1991,Giamarchi1995,gingras1996,Okamoto2015} in a CDW system.  
{\it v}) None-the-less, vestiges of the two distinct CDW phases persist all the way to 5\% Pd-intercalation, far beyond where signatures of the smeared phase transitions are observed in bulk  probes.

\begin{figure}[h]
\includegraphics[width=1.0\columnwidth]{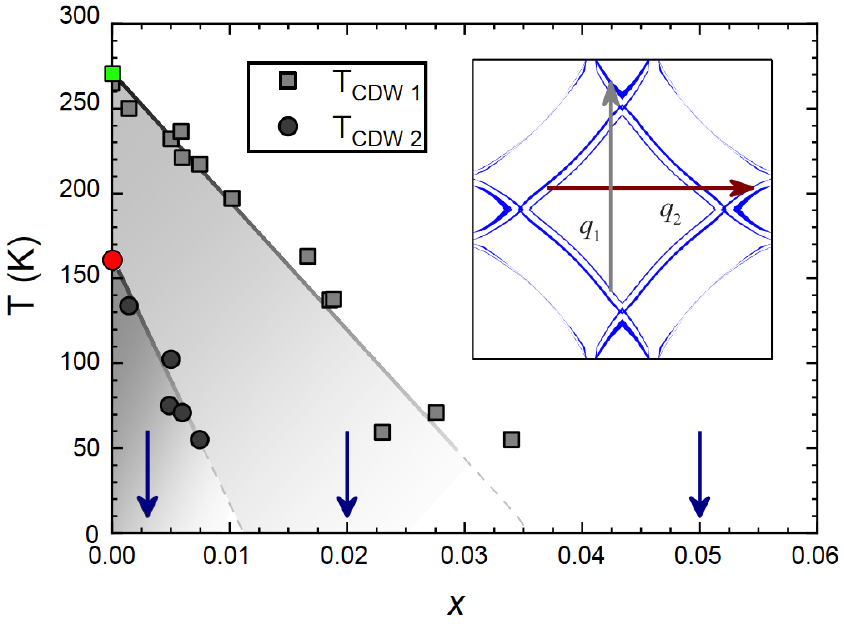}
\caption{Phase diagram taken from \cite{Straquadine2019}. Colored points at $x=0$ mark the true phase transitions of the pristine sample. Grey points at finite $x$ mark the crossover temperature below which finite-range CDW order appears as features in the electrical resistivity. Vertical arrows mark the three Pd concentrations discussed in this manuscript. Inset shows calculated Fermi surface of the parent compound ErTe$_3$ seen down the long $b$-axis (reproduced from \cite{Ru2006}). }
\label{basic}
\end{figure}
 
ErTe$_3$ belongs to the RTe$_3$  (R = rare-earth) family of quasi-2D metals which exhibit unidirectional incommensurate charge density wave states \cite{DiMasi1995,Ru2006,Fang2007,Ru2008}. It is formed in the orthorhombic space group $C_{mcm}$ \cite{Norling1966}, and contains double layers of nominally square Te planes separated by RTe block layers. Orthorhombicity is derived from a glide plane stacking of these tetragonal layers which determine the electronic properties of this system at high temperatures. A unidirectional CDW was first detected in this system by transmission electron microscopy (TEM) \cite{DiMasi1995}. Further studies of angle resolved photoemission spectroscopy (ARPES) have shown that large portions of the Fermi surface (FS) nested by ${\bf q}_{CDW}$ are indeed gapped \cite{Brouet2004}. Scanning tunneling microscopy (STM) and spectroscopy (STS)  established the fully incommensurate nature of the CDW in TbTe$_3$, while also revealing an additional CDW ordering perpendicular to the principal one \cite{Fang2007}, but weaker in amplitude. Subsequent ARPES studies on ErTe$_3$ established that two incommensurate CDW gaps are created in two step transitions by perpendicular FS nesting vectors \cite{Moore2010}. Despite the near-nesting conditions that result in gapping of a substantial portion of the Fermi surface in the CDW state, several factors indicate that strong coupling effects unrelated to Fermi surface nesting play a role \cite{Johannes2008}. Some evidence for such a perspective might be found in the large values of $2\Delta_{CDW}/k_BT_{CDW}$ ($\sim15$ and $\sim7$ for the first and second CDW respectively)  \cite{Varma1983}, though several other factors can also yield such an effect. More convincingly, high-energy-resolution inelastic x-ray scattering investigation of TbTe$_3$ revealed strong phonon softening and increased phonon linewidths over a large part in reciprocal space adjacent to the CDW ordering vector, thus showing momentum-dependent electron-phonon coupling \cite{Maschek2015}. Resistivity measurements resolved in the two perpendicular in-plane directions in pristine ErTe$_3$ show a kink at each of the CDW transitions, and it was observed that the introduction of Pd intercalation broadens and suppresses both features to lower temperatures (see Fig.~\ref{basic}).\\

\noindent {\bf Experiment} - 
Pd$_x$ErTe$_3$ with $0\leq x \leq 0.055$ were grown using a Te self-flux as described elsewhere for pure RTe$_3$ compounds \cite{Ru2006}, with the addition of small amounts of Pd to the melt \cite{Straquadine2019}. The resulting Pd concentration was determined by electron microprobe analysis.
Three levels of intercalation were studied: 0.3\%, 2\% and 5\% (marked in Fig.~\ref{basic}). Prior to STM measurements, samples were characterized as described in \cite{Straquadine2019}.
STM was performed with a hybrid UNISOKU-USM1300 system constructed with a home-made ultra high vacuum sample preparation and manipulation system. The samples were cleaved  at room temperature at pressures of low $10^{-10}$ torr and immediately transferred to the low temperature STM.  Topography was performed at $\approx1.7$ K, with typical tunneling parameters of $V_{bias}$=50-100 mV and I=100-300 pA.\\

\noindent {\bf Results and Discussion} - 
The phase diagram in Fig.~\ref{basic} shows that the weaker CDW state along the $a$-axis is more sensitive to disorder.  Its ``transition temperature'' is suppressed quickly as $x$ increases, and extrapolates to zero around $x = 1\%$. Thus,  samples with low levels of intercalants  should be an excellent starting point to observe the effect of weak disorder on both transitions.  
\begin{figure}[h]
\includegraphics[width=1.0\columnwidth]{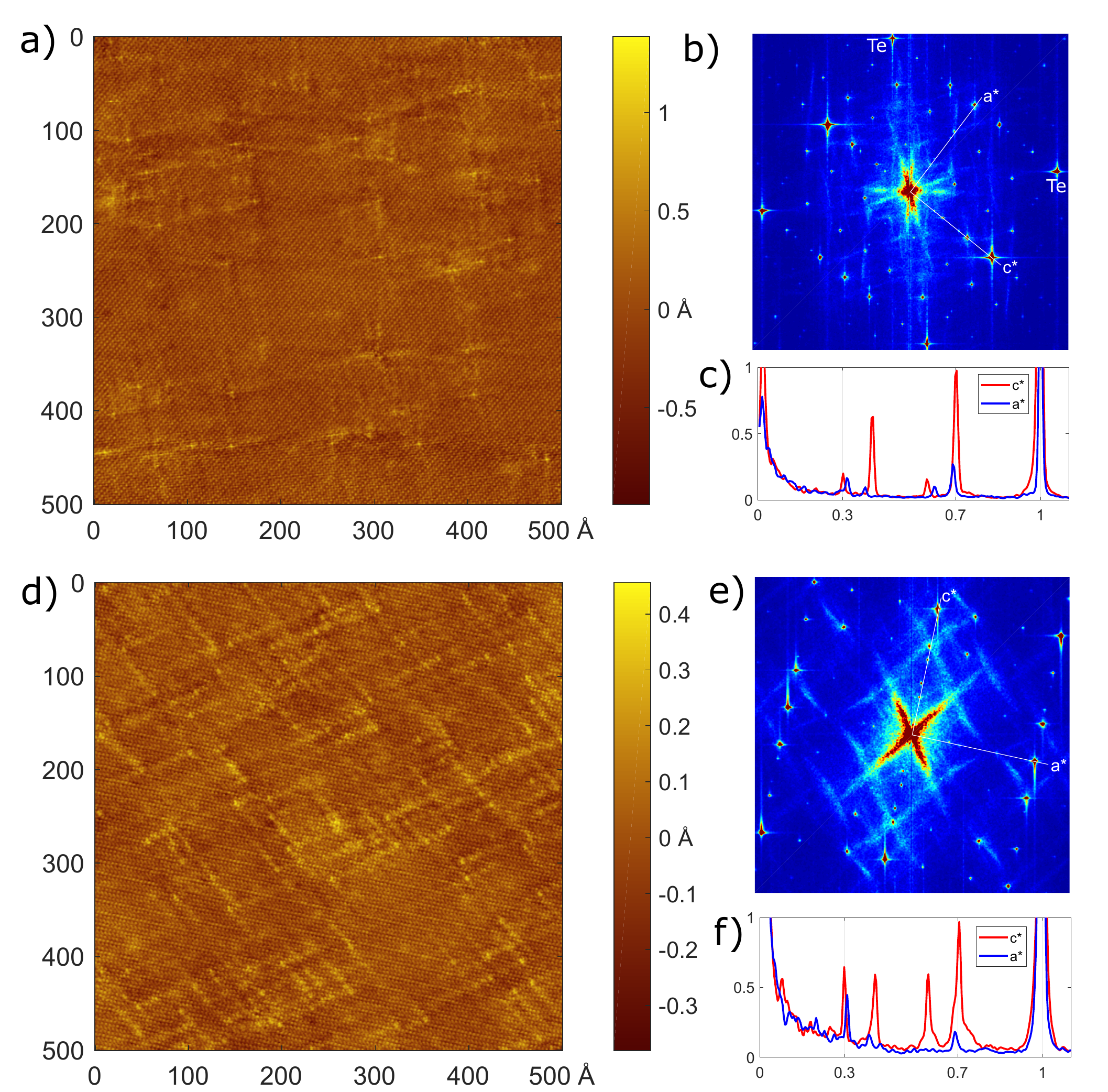}
\caption{a)Topography of 0.3\% Pd-intercalation sample. b) FFT of (a) with lattice points from the Te plane labeled and white lines indicating the line cut directions.   c) The respective line cuts.   Here stronger CDW peaks (red) are along $c^*$. d)Topography of 2\% Pd-intercalation sample. e) FFT of (d) and f) the respective line cuts.}
\label{figx}
\end{figure}
Fig.~\ref{figx}a shows a topography of a $x = 0.3\%$ sample.  The features visible are an atomic corrugation, CDW corrugations, as well as small lumps on the surface, often with a skewed ``$\times$'' pattern around them.  The lumps, which with simple counting roughly agree with the intercalation density, are only a fraction of an angstrom high, and thus must arise from sub-surface intercalants rather than surface Pd atoms. Single-crystal x-ray diffraction on the same crystals confirmed the increase in the $b$-axis lattice constant consistent with Pd atoms intercalating between the van der Waals bonded Te bilayer \cite{Straquadine2019}.  DFT calculations also show this to be the most favorable location for intercalation \cite{He2016}.  No surface Pd atoms were observed for low intercalations, presumably due to the high volatility of the Pd atoms upon cleaving at room temperature. 

Much of the analysis of the CDW and effects of disorder, are done via the Fourier transform (FFT) of the topographic data, as in Fig.~\ref{figx}b. The dominant CDW generates a series of very sharp peaks from the origin to both lattice points corresponding to the RTe block layer crystal structure. As in previous STM and X-ray studies of the parent compound \cite{Fang2007,Yao2006,Ru2006}, there are four characteristic CDW points due to mixing with the lattice of $q_{CDW1} =0.70c^*$, $1-q_{CDW1} = 0.30c^*$, $2-2q_{CDW1} =  0.60c^*$ and $2q_{CDW1}-1 = 0.40c^*$.   (In this material, the four-fold symmetry is broken by a glide plane, rendering the structure weakly orthorhombic; in the standard setting for the orthorhombic space group, the wavevector of the primary CDW lies along the c-axis.) 
To identify and distinguish the two perpendicular CDWs, we take line cuts in the FFT data from the origin to either the a* (blue) or c* (red) lattice points. (Fig.~\ref{figx}c)  We see a set of peaks with $q_{CDW1} =0.70c^*$, $0.30c^*$, etc., and another set with $q_{CDW2} =0.68a^*$, $0.32a^*$, etc.  (The uncertainty is approximately $0.005c^*$, or one FFT pixel, in our 100 nm scans.)  Based on the wavelength alone, we can identify the former with the ``first", or higher Tc, CDW which forms in the c-axis and breaks the nearly four-fold symmetry of the material  \cite{Ru2008,Brouet2008,Moore2010}.  The latter set of peaks is the ``second" CDW transition which occurs at lower temperatures, and is the one that shows a weaker signal in diffraction measurements and disappears first with rising temperature or intercalation level \cite{Straquadine2019}.  In our STM data, depending on many factors such as tunneling parameters, tip condition, and scan location, the relative amplitudes of the four peaks within one line scan can vary. However, it is generally true that the CDW1 signal is always stronger everywhere in the $0.3\%$ sample, thus giving the visual impression of a unidirectional CDW.
We also note the presence of strong satellite peaks at $\textit{\textbf{q}}_{Te} \pm \textit{\textbf{q}}_{CDW}$.  The CDW exists in the Te plane, and its signal is strongly modulated by this lattice ($\textbf{q}_{Te}$).  Similar strong satellites were also seen in STM data of pristine TbTe$_3$ \cite{Fang2007}.

Next we present the effects of a higher level of intercalation, where the secondary CDW is expected (from resistivity data) to be absent.  Data for $2\%$ intercalation is shown in Fig.~\ref{figx}d-f. Similar to the $0.3\%$, we note the lack of  surface Pd atoms. There also exists the same sub-angstrom lumps, the ``$\times$'' pattern, and their greater abundance is in accordance with with the higher intercalation level. However, even at this intercalation level, signatures of both CDWs are present, with the second CDW having a lower amplitude, as also seen in the low temperature X-ray data of \cite{Straquadine2019}

Unlike the parent compound, the presence of intercalants creates additional features in the topography and FFT due to quasiparticle scattering interference (QPI).  In real space, this manifests itself as the streaks that emanate from some of the ``lumps" in a skewed ``$\times$'' pattern.  In Fourier space, it manifests as a compressed ``$\times$'' at the origin, which is biased in the c* direction in the $2\%$ sample, and may reflect the broken 4-fold symmetry via two different CDW gaps \cite{Moore2010}.  In the $0.3\%$ sample there are two of these features, perpendicularly overlapped. This ``$\times$'', as well as the other streaks in the FFT most likely come from quasiparticle scattering along un-gapped portions of the Fermi surface, which generally run $\approx$ 45 degrees to the CDW vectors.  (See Fig.~\ref{basic})

\begin{figure}[h]
\includegraphics[width=1.0\columnwidth]{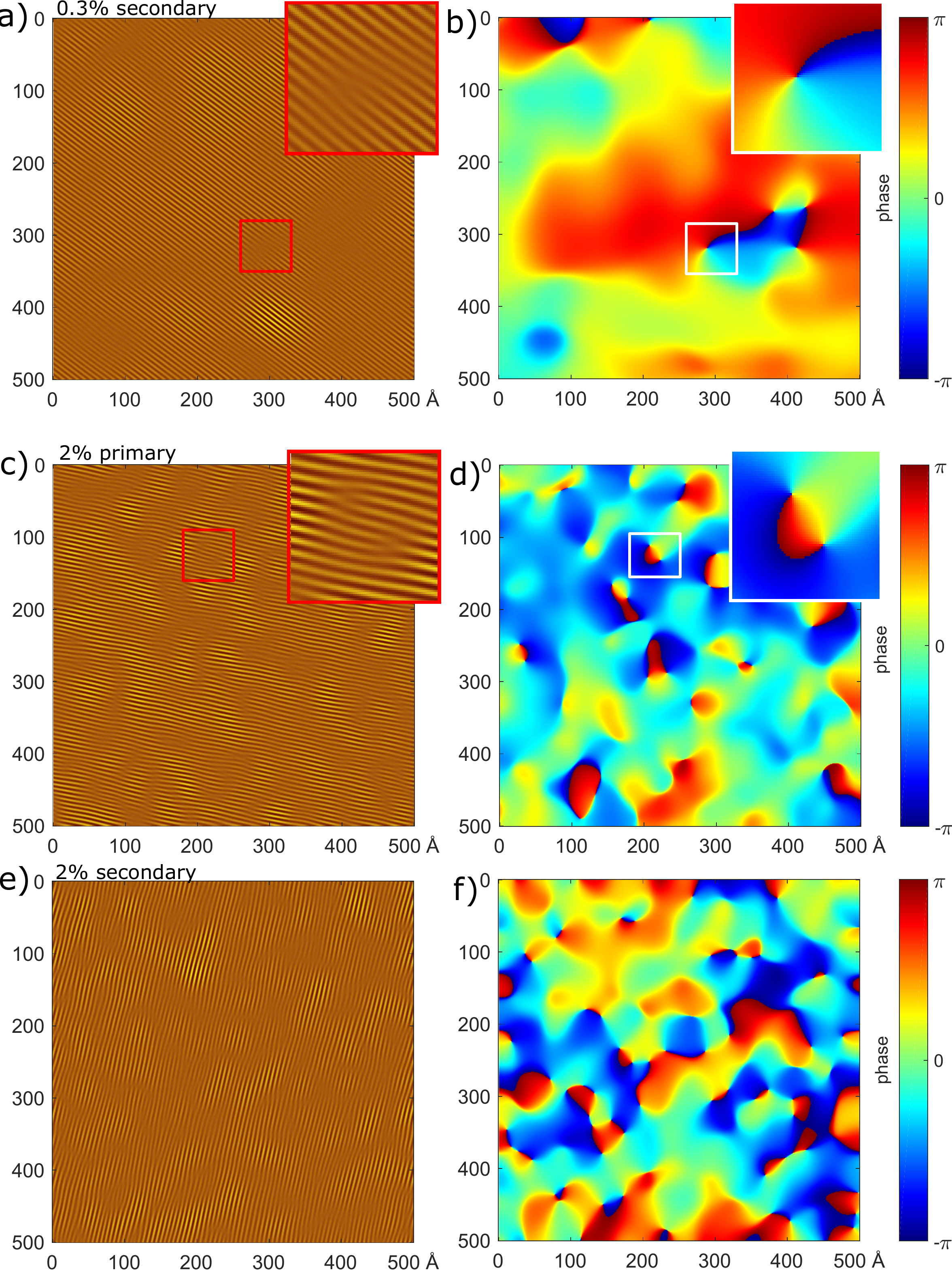}
\caption{CDW defect analysis of 0.3\% and 2\% Pd-intercalation samples. a) Topography of Fig.~\ref{figx}a) (0.3\%) filtered around the secondary (weaker) 0.68a* CDW point and highlighting a dislocation. b) Phase. Similar analysis of the primary (stronger) CDW phase along $c^*$ yields no observed dislocations and is thus not shown. c),e) Topographies of the 2\% sample filtered around the primary and secondary CDW points, respectively.  d,f) Phase of the primary and secondary CDW, respectively. Inset shows a paired dislocation.}
\label{filterphase}
\end{figure}

To analyze the effect of the impurities on the CDW, we filter the data in Figs.~\ref{figx}a,d around the $q_{CDW1} = 0.70c^*$ (or $0.68a^*$) point and plot it in Figs.~\ref{filterphase}a,c,e.  Highlighted in the red rectangle, and further zoomed in the upper corner, is a dislocation (with no neighbors within $\gtrsim$ 100 \AA) for the $0.3\%$ sample, and a pair of dislocations for the $2\%$ sample.  Two CDW stripes merge into one, and within a few tens of angstroms, the opposite occurs.  The result is that outside of this region, the two dislocations cancel out and the total number of CDW stripes is conserved.  The effect is further quantified when we do a variant of the Fujita-Lawler analysis of a localized Fourier transform in order to get the phase of the CDW \cite{Lawler2010,Fang2018}. This is shown in Figs.~\ref{filterphase}b,d,f.  Taking a closed-path around each of the dislocations, the phase picks up 2$\pi$, but outside the pair, the phase is continuous.  In the $0.3\%$ sample, there are few dislocations within the (500 \AA)$^2$ scan range for the secondary CDW (Fig.~\ref{filterphase}b), and typically no dislocations for the primary CDW (shown in the Supplementary \cite{supplemental}). As explained in the theoretical discussion in the Supplemental Materials, this is consistent with it being in a stripe-glass phase, which is a variant of the Bragg glass phase \cite{Giamarchi1995}. For the $2\%$ sample, there are overall more dislocations (Figs.~\ref{filterphase}d,f), with the secondary CDW having slightly more than the primary (indicating a shorter correlation length).  (Significantly, the dislocations in the two perpendicular directions appear uncorrelated.)  
 
We see phase variations over length scales of tens of \AA\ in the $0.3\%$ and $2\%$ samples, which represent mild distortions in the CDW from the impurities (even if there is no dislocation).  Within our scan ranges, there are no domains of exclusively one CDW direction over the other, which rules out unidirectional (stripe) phases in favor of bidirectional.
 
\begin{figure}[h]
\includegraphics[width=1.0\columnwidth]{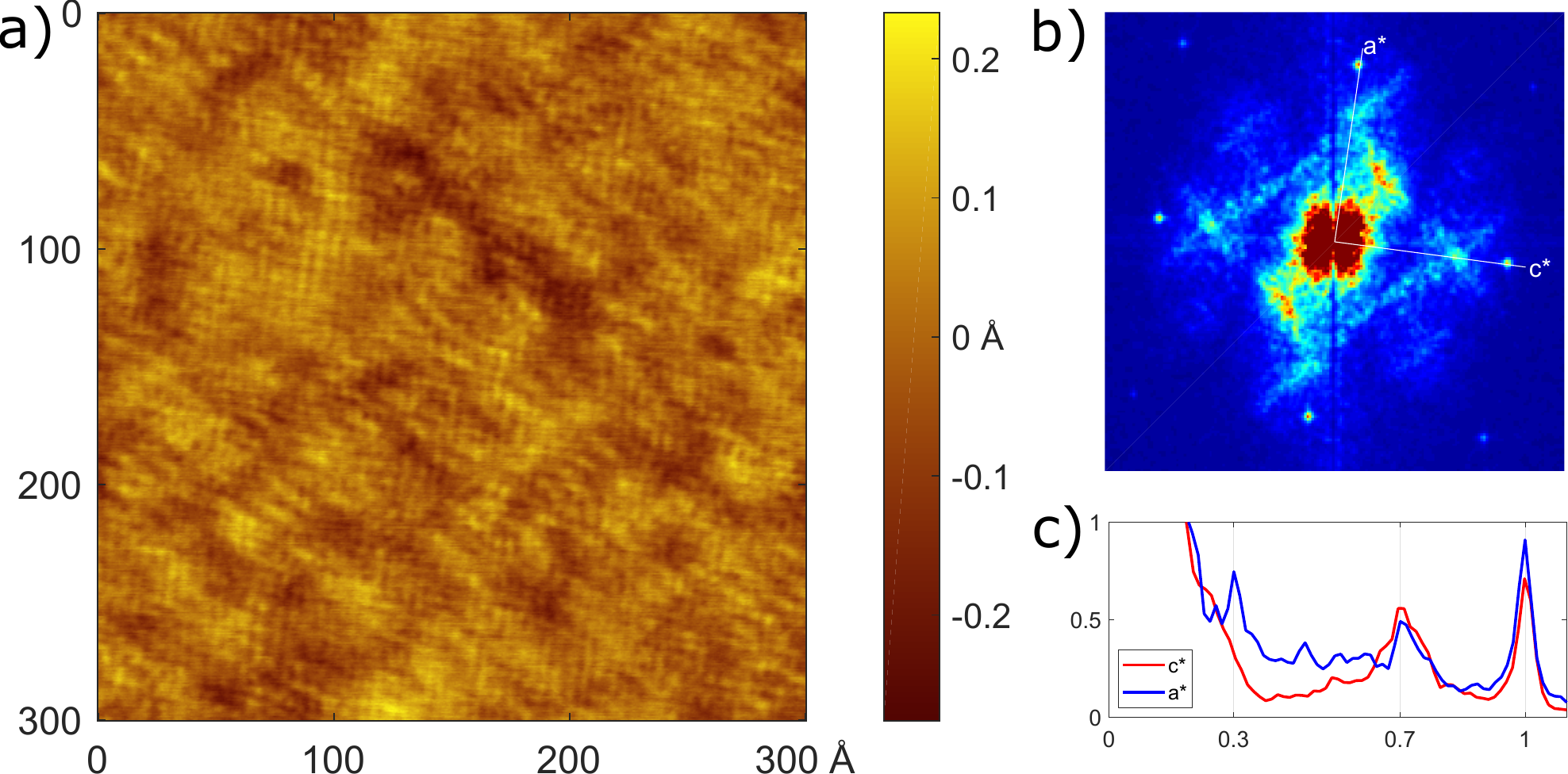}
\caption{a)Topography of 5\% Pd-intercalation sample. b) FFT of (a) and c) the respective line cuts. }
\label{fivepercent}
\end{figure}

Finally, we discuss the $5\%$ Pd-intercalation sample, where  Fig.~\ref{basic}  would suggest that remnants of both CDW phase transitions are completely suppressed. Fig.~\ref{fivepercent}a shows a topographic image of the $5\%$ Pd-intercalation sample taken at 1.7 K.  It is qualitatively different than the lower intercalation samples, lacking the skewed  ``$\times$'' pattern, but still having sub-\AA\ height modulations with small length-scale corrugations.
We typically observe an adatom on the surface approximately once every 30 nm, of height a few \AA, which occasionally moves or causes a tip reconstruction.  This limits our scan size and therefore FFT resolution. 

Examination of the FFT (Fig.~\ref{fivepercent}b) reveals that the typical CDW peaks that appeared at lower Pd concentrations are now significantly broadened, as shown in the line cuts (Fig.~\ref{fivepercent}c).  In particular, no sharp $k$-space points were seen in X-ray and electron diffraction measurements at intermediate temperatures (95 K) \cite{Straquadine2019}.   Instead, broad and diffuse streaks were seen spanning between the original CDW points, indicating 2 dimensional short range correlations that are consistent with the q-dependent susceptibility. What could be these same streaks appear in our data  between the CDW points as well.  In fact, the visible corrugations seen in the topograpahy are due to these streaks plus the broad CDW peaks, and not the atomic lattice, although the lattice points are well-defined in the FFT.
 This short range CDW order likely originates from CDW fluctuations pinned by the disorder as the sample is cooled to the STM working temperature \cite{KivRMP2003}. A similar observation was reported for the CDW system NbSe$_2$ \cite{Arguello2014}\\

\noindent {\bf Conclusions} - 
Large area STM scans of Pd-intercalated ErTe$_3$ allow for a direct comparison with diffraction data. High spatial resolution local measurements probe the origin of the destruction of the two distinct, mutually perpendicular, incommensurate CDW orders that coexist in the parent material.  The Pd intercalants, which are local on the atomic scale, create a larger concentration of dislocations in the secondary CDW.   Proliferation of dislocations is the driving mechanism by which the long range order of the CDW state is destroyed.   Even without CDW order, there is a large degree of orientational long-range (nematic) order.  Upon increased levels of intercalation and thus increased disorder, only very short range CDW order persists. Finally, the fact that at the lowest disorder levels, we observe dislocations in CDW2 but not CDW1 is suggestive of the existence of a nematic glass phase - a specific version of a ``Bragg glass.'' 
\bigskip

\noindent {\bf Acknowledgments}:
  This work was supported by  the U.~S.~Department of Energy (DOE) Office of Basic Energy Science, Division of Materials Science and Engineering at Stanford under contract No.~DE-AC02-76SF00515. JS is supported by an ABB Stanford Graduate Fellowship. Various parts of the STM system were constructed with support from the Army Research Office (ARO), grant No.~W911NF-12-1-0537 and by the Gordon and Betty Moore Foundation through Emergent Phenomena in Quantum Systems (EPiQS) Initiative Grant GBMF4529. 

\bibliography{pdxerte3}
\newpage

\setcounter{figure}{0}
\renewcommand{\thefigure}{S\arabic{figure}}%

{\centering 

{\large Supplemental Material for:}
\bigskip

{\bf Disorder Induced Suppression of CDW Long Range Order: STM Study of Pd-intercalated ErTe$_3$}\\
{\normalsize Alan Fang,$^{1,2,3}$ Joshua A. W. Straquadine,$^{2,3}$ Ian R. Fisher,$^{1,2,3}$ Steven A. Kivelson,$^{1,2,4}$ and Aharon Kapitulnik$^{1,2,3,4}$\\
}}

\section{ I. MATERIALS AND EXPERIMENTAL METHODS}
In this section we provide details on the experimental methods and the analyses of the experimental data. 
\vspace{-3mm}

\subsection{Single Crystals Growth and Characterization}
\vspace{-3mm}
Pd$_x$ErTe$_3$ with $0\leq x \leq 0.055$ were grown using a Te self-flux as described elsewhere for pure RTe$_3$ compounds \cite{Ru2006}, with the addition of small amounts of Pd to the melt \cite{Straquadine2019}. Plate-shaped crystals ($b$-axis normal to the plane) 1-3 mm across were routinely produced. Microprobe chemical analysis was used to determine the amount of Pd incorporation into the samples, while the decrease in crystals thickness from several hundred microns for $x = 0$ to approximately 50 microns for x = 0.05 served as evidence that Pd atoms act as intercalants between the weakly-coupled Te planes. Resistivity and magnetic susceptibility were used to track the transition temperature and sharpness of the two unidirectional CDW phases, and the emergence of superconductivity with increasing intercalation level. Single-crystal x-ray diffraction was used to determine the lattice parameters, and in particular the relation between intercalation and the increase in the $b$-axis lattice constant as a result of the Pd intercalation. The $a$- and $c$ in-plane lattice parameters are unchanged within experimental error, arguing against chemical substitution of Pd on the Er site, as the smaller radius of Pd ions would be expected to compress the lattice in-plane.  A detailed description of sample preparation, characterization and the effect of Pd intercalation on the bulk properties are given in \cite{Straquadine2019}.

\subsection{Scanning Tunneling Microscopy and Spectroscopy Measurements}
\noindent {\bf The apparatus -} 

Scanning tunneling microscopy (STM) was performed with a hybrid UNISOKU USM1300 system (base temperature 370 mK) plus home-made Ultra High Vacuum (UHV) sample preparation and manipulation system. The samples were cleaved  at room temperature and at pressures of low $10^{-10}$ torr and immediately transferred to the low temperature STM.  Topography was performed at ~1.7 K with typical tunneling parameters of $V_{bias}$=50-100 mV, I=100-300 pA, and scan sizes up to 100 nm with atomic resolution.

\noindent {\bf Additional Data on Analysis of Dislocations-} 

\begin{figure}[h]
\includegraphics[width=1.0\columnwidth]{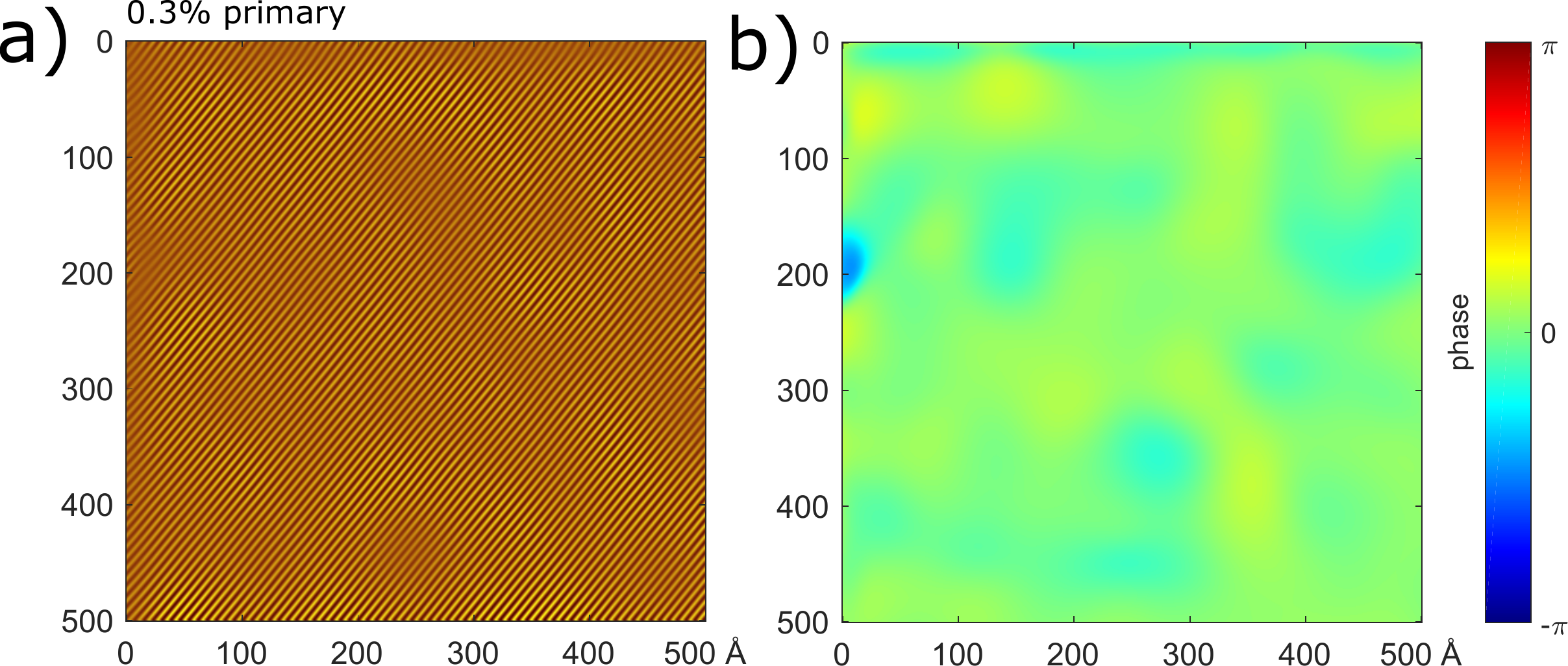}
\caption{Supplementary figure to the CDW defect analysis of 0.3\% Pd-intercalation sample, using the topographic data in Fig.~\ref{figx}a. a) Filtered CDW b) Phase.}
\label{zp3_stronger}
\end{figure}

As explained in the main text, to analyze the effect of the impurities on the CDW, we filter the topographic data around the $q_{CDW} = 0.70c^*$ (or $0.68a^*$) point to obtain a real-space representation of the CDW.  To highlight distortion and dislocations,  we employ a variant of the Fujita-Lawler analysis of a localized Fourier transform in order to get the phase of the CDW \cite{Lawler2010,Fang2018}. Not shown in the main body of the paper is this analysis for the 0.3\% 
 primary CDW, which lacks dislocations.  (Fig.~\ref{zp3_stronger}) Note the un-interrupted CDW, and the fairly uniform phase.   (It is possible that a dislocation exists beyond the left of the scan area)

\section{II. Theoretical Background  }
In this section we provide a summary of the theoretical background behind our analyses of the experimental data. It sketches an effective field theory of two component CDW order in the presence of quenched randomness. In reality, the crystals involved are slightly orthorhombic, but this seems to be a very small effect so we will consider an idealization in which the crystal is taken to be tetragonal and the orthorhombicity -- when included at all -- will be represented as an infinitesimal external symmetry breaking field. 
\vspace{-3mm}

\subsection{Reviewing the phase diagram in the clean limit}
\vspace{-3mm}
 In the absence of disorder (or the best achievable approximation  to this limit) the phase diagram in the $T$ and chemical pressure $P$ shows three distinct phases:  a uniform (no CDW) phase for $T> T_{CDW1}(P)$, a phase with ``bidirectional'' CDW order, but with distinctly different strengths and slightly different ordering vectors for the two CDW components at $T < T_{CDW2}(P) < T_{CDW1}(P)$, and a ``unidirectional'' CDW for intermediate temperatures $T_{CDW1}(P) > T > T_{CDW2}(P)$.  Both CDW phases break the tetragonal point-group symmetry, and in this sense have a nematic component. 

Since the two transitions are separated from each other, it is reasonable to treat them separately.  However, since the various ordering tendencies get both truncated and scrambled when disorder is introduced, we would like to treat them from the perspective of a single, unified effective field theory.  Moreover, it is observed that as a function of increasing $P$, the two lines tend to approach one another, so that for the case of ErTe$_3$ studied here (which has the largest effective $P$ - i.e. the smallest lattice constant of any of the stoichiometric  tri-tellurides studied to date) $(T_{CDW1}-T_{CDW2})/T_{CDW1} \approx 0.4$.  We would thus like to think that this allows us to organizing our thinking about a putative multicritical point at $T=T^*>0$ and $P=P^*$ at which these two lines would meet at slightly higher pressure, i.e. $T_{CDW1}(P^*)=T_{CDW2}(P^*) = T^*$.

The nature of such a multicritical point in a tetragonal system is already somewhat unusual.   
To see this, consider the lowest order effective (Landau) potential as a function of  the two components of the CDW order represented by two complex scalar fields, $\phi_1$ and $\phi_2$.  To the extent that the crystal can be treated as approximately tetragonal, the effective Hamiltonian density, ${\cal H}[\phi_1,\phi_2]$, is symmetric under discrete rotations (and other point group interactions) that exchange $\phi_1$ and $\phi_2$.  (The weak effect of the subtle orthorhombicity of the actual crystal structure can be modeled as a small symmetry breaking term of the form ${\cal V}_{orth}\equiv -b[ |\phi_1|^2-|\phi_2|^2 ]$.)  

\subsection{$\phi^4$ theory of stripe and nematic phases}
\vspace{-3mm}
To fourth order in the fields, the effective potential is 
\bea
{\cal V}(\phi_1,\phi_2)=&& \frac \alpha 2 \left[|\phi_1|^2 + |\phi_2|^2\right] \\
&&+ \frac 14  \left[|\phi_1|^2 + |\phi_2|^2\right]^2 + \frac \gamma 2 |\phi_1|^2|\phi_2|^2 + \ldots \nonumber
\eea
where we have assumed that the quartic term is positive and have normalized the fields such that its strength is unity.  

\begin{figure}[h]
\includegraphics[width=1.0\columnwidth]{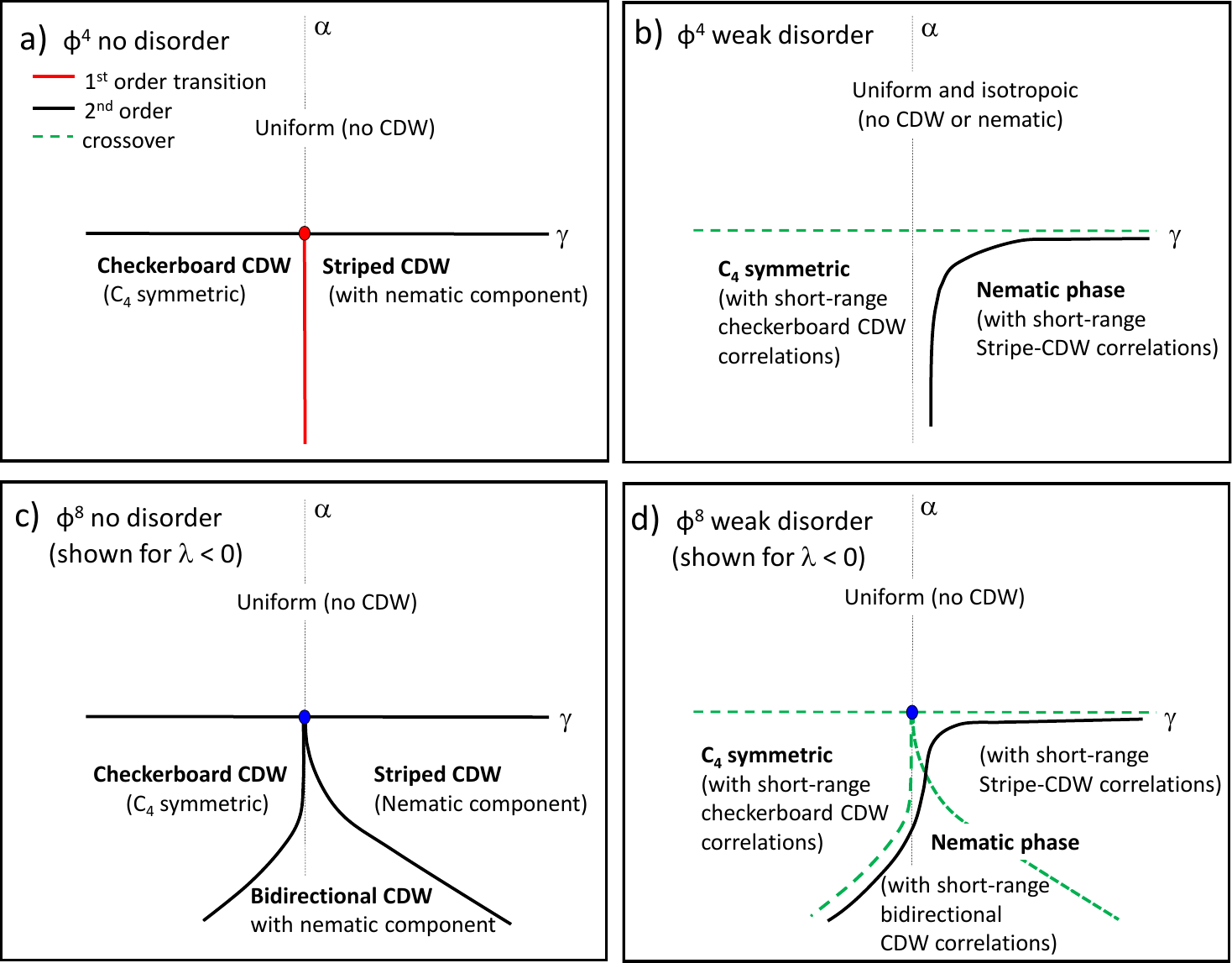}
\caption{Steps in constructing the phase diagram which is the result of the minimization of the different parts of the effective potential - see text.}
\label{theory1}
\end{figure}

The phase diagram in the $\alpha-\gamma$ plane that results from minimizing this effective potential (i.e. Landau theory) is shown in Fig.~\ref{theory1}a.  For $\gamma >0$, there is a second-order transition to a stripe ordered CDW phase as $\alpha$ changes from positive to negative values, while for $\gamma <0$ the transition is to a checkerboard state.  In the stripe phase, either the thermal average of $\phi_1$ is non-zero and $\phi_2=0$ or the converse;  in addition to breaking translational symmetry in one direction, this phase strongly breaks the $C_4$ rotational symmetry that interchanges the two components of the order parameter.  In the checkboard phase, the thermal averages of $\phi_1$ and $\phi_2$ are not only non-zero, but of equal magnitude;  this state breaks translation symmetry in both directions but leaves an unbroken $C_4$ rotational symmetry.  The point $\alpha=\gamma =0$ is a bicritical point, below which, as a function of $\gamma$, there is a first order boundary between the stripe and checkerboard phases.  Fluctuation effects for an effective Landau-Ginzburg-Wilson model with this form of the effective potential have been considered elsewhere - see especially in~\cite{Nie2014}.  In addition to their effects on critical exponents, these can lead to additional subtleties in the nature of the phase diagram and even to additional phases making their appearances;  for instance, in a quasi-2D system, there can be a narrow strip between the stripe ordered phase and the fully symmetric phase in which CDW order is melted but in which vestigial $C_4$ symmetry breaking persists, giving rise to a vestigial nematic phase. 

More interesting is what happens to this phase diagram in the presence of weak but non-negligible disorder.  Here (in $d \leq 4$) no incommensurate CDW long-range order is possible, which likely implies that the CDW transitions are all replaced by crossover lines separating regions with little in the way of CDW correlations from regions with substantial intermediate range CDW order.  (An interesting possibility, which we will ignore for now but to which we will return below, is that for weak enough disorder there could exist a Bragg glass phase with power-law CDW order and no unbound dislocations \cite{Giamarchi1995,gingras1996}.)  However, as shown in Ref.~\cite{Nie2014}(and as would anyway be expected on general grounds) the nematic component of the stripe order persists up to a non-vanishing critical disorder strength.  The shape of the phase boundary follows from continuity - in the limit that the disorder strength tends to zero, this boundary must coincide with the phase boundary of whatever CDW phase has a nematic component.  Thus, the phase diagram in the presence of weak disorder becomes that in Fig.~\ref{theory1}b, where the dashed lines indicate crossovers and the solid black line bounds the nematic phase.

\subsection{$\phi^8$ theory and the nematic bidirectional CDW phase}
\vspace{-3mm}
Manifestly, the $\phi^4$ theory is inadequate, in that it is missing one phase that is observed in experiments (the nematic bidirectional CDW), and yet has one implied phase (the checkerboard CDW) not seen in experiments.  While we can always imagine that the checkerboard CDW is lurking, as yet undiscovered, in the large $P$ reaches of the phase diagram, the missing bidirectional phase is something that needs to be addressed.  

Even in the case of pure ErTe$_3$, $T_{CDW2}$ is smaller than $T_{CDW1}$ by a substantial factor, so at temperatures in the neighborhood of $T_{CDW2}$ there is no justification for assuming that $|\phi_j|$ is small; there is thus no reason to keep only low order terms in powers of the field in the effective field theory.  Indeed, away from the putative multicritical point, it is legitimate to treat the two clean-limit transitions in terms of two distinct effective field theories.  Let us consider the case where we are well below $T_{CDW1}$, where the value of either $|\phi_1|$ or $|\phi_2|$ is not small, but above $T_{CDW2}$, and even slightly below it, the subdominant component is a legitimate expansion parameter.  It is convenient to carry out the expansion of the effective potential in a way that is still manifestly $C_4$ invariant.  We thus define new quantities as follows:  $\phi_1 \equiv  |\phi| \cos(\theta)e^{i\delta_1}$ and $\phi_2 \equiv  |\phi| \sin(\theta)e^{i\delta_2}$.  Translational symmetry implies that the free energy must be independent of $\delta_1$ and $\delta_2$, and the tetragonal symmetry implies that the free energy must be invariant under $\theta \to -\theta$ and $\theta \to \theta + \pi/2$.  We can therefore in complete generality express the effective potential in terms of the two quantities $|\phi|^2\equiv |\phi_1|^2 + |\phi_2|^2$ and $\Delta^4 \equiv |\phi_1|^2|\phi_2|^2= |\phi|^4[1-\cos(4\theta)]/8$, such that near $T_{CDW2}$, $\Delta$ is small.  We thus expand ${\cal V}$ in powers of $\Delta$ giving
\be
{\cal V}= {\cal V}_0(|\phi|^2) + \frac{{\cal V}_1(|\phi|^2)}2 \ \Delta^4+\frac { {\cal V}_2(|\phi|^2)}4 \Delta^8 + \ldots
\ee
where (noting explicitly for future use all terms of order up to order $|\phi|^8$ and using the same conventions for the low order terms as above)
\be
{\cal V}_0(\phi^2) = \frac {\alpha}2 \phi^2 + \frac 1 4 \phi^4 +\frac {u_6} 6 \phi^6+\frac {u_8} 8 \phi^8+\ldots
\ee
\be
{\cal V}_1(\phi^2) = {\gamma}  + \frac {\gamma_2} 2 \phi^2 + \frac {\gamma_4} 4 \phi^4\ldots
\ee
\be
{\cal V}_2(\phi^2) =  {\lambda}  + \ldots
\ee

In the neighborhood of $T_{CDW2}$, $|\phi|$ is already large and does not change much in magnitude, so we can focus exclusively on $\theta$.  For $V_1 > |V_2||\phi|^4/4$, ${\cal V}$ is minimal for $\theta = 0$  {\it i.e.} we are in the stripe phase, while for $V_1 < - |V_2||\phi|^4/4$ the minimum is at $\theta =\pi/4$  {\it i.e.} we are in the checkerboard phase.    (Here $[\theta]$ means the value of $\theta$ mod $\pi/2$ chosen to lie in the interval $0\leq [\theta] < \pi/2$.) The shape of this intermediate regime depends on the sign of $V_2$.  For $V_2 <0$ (the case illustrated in Fig.~\ref{theory1}c), in the  interval $0 < V_1 < |V_2||\phi|^4$,
\be
\theta= \frac 1 4 \cos^{-1} \left[\frac {|V_2||\phi|^4-8V_1}{|V_2||\phi|^4}\right].
\ee

 Thus, the boundary between the stripe phase and the nematic bidirectional CDW  phase ({\it i.e.} the line that implicitly defines $T_{CDW2}$) occurs where $V_1=4|V_2|\phi^4$, while the boundary between the nematic bidirectional CDW  phase and the checkerboard phase occurs where $V_1=0$.  
On the other hand, for  $V_2>0$, the stripe phase persists  as long as $V_1>0$.  Thus, $T_{CDW2}$ is defined by the condition $V_1=0$, while the transition to the checkerboard phase occurs where $V_1=-V_2|\phi|^4/4$.  In both cases,  all transitions are continuous.

 For simplicity, this discussion was carried through in the case in which $T_{CDW2}$ is well separated from $T_{CDW1}$.  However, clearly the same considerations apply even as they approach each other at the multicritical point, $\alpha=\gamma=0$.  All that is different here is that the size of the intermediate nematic bidirectional CDW phase, which is bounded by $|\delta {\cal V}_1|< |{\cal V}_2|\phi^4/4$, and thus gets to be parametrically small, as shown in Fig.~\ref{theory1}c.  Here, approximating these terms by their leading order expressions in powers of $|\phi|^2$, we infer that the  intermediate phase occurs for a range of $\gamma$ such that $|\gamma| \lesssim |\lambda||\alpha|^2$.

It is interesting to note that the instability of the bicritical point seen in the  $\phi^4$ treatment and its substitution by this peculiar tetracritical point appears to be inevitable - at least at the level of mean-field theory.  We have not explicitly carried out the analogous treatment of  fluctuation effects and the effects of disorder in the resulting $\phi^8$ theory as was carried out for the $\phi^4$ theory.  Away from the multicritical point, however, more or less the same considerations apply, leading to the (conjectural) phase diagram shown in Fig.~\ref{theory1}d in the presence of weak disorder.

\begin{figure}[h]
\includegraphics[width=0.9\columnwidth]{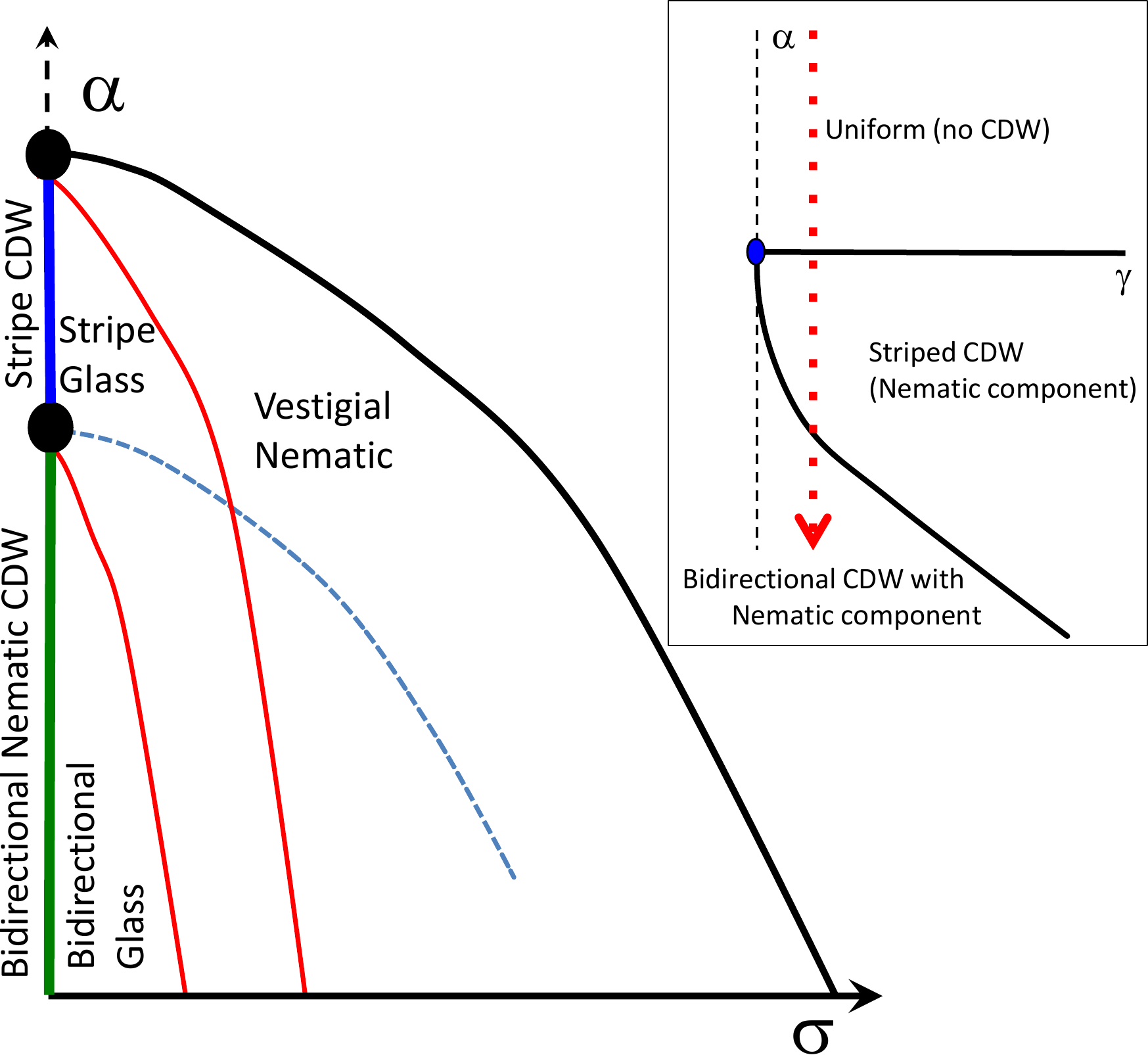}
\caption{A conjectured phase diagram as a function of temperature $T$ and disorder strength $\sigma$. Here, the inset shows a portion of the clean limit phase diagram (plotted in the $\alpha-\gamma$ plane, where $\alpha$ should be considered a proxy for $T$) - see text}
\label{theory2}
\end{figure}

Indeed, at the conjectural level, it is reasonable to propose that as a function of $T$ and disorder strength $\sigma$, one would expect a phase diagram with the general topology shown in Fig.~\ref{theory2}.  Here, the inset shows a portion of the clean limit phase diagram (plotted in the $\alpha-\gamma$ plane, where $\alpha$ should be considered a proxy for $T$) and the dotted red line shows a trajectory through this diagram at fixed $\gamma$ such that there are the two requisite transitions at $T_{CDW1}$ and $T_{CDW2}$.  The main figure assumes the same fixed $\gamma$ but shows a putative phase diagram in the $\alpha-\sigma$ plane.  Phases with CDW long-range order are confined to the $\sigma=0$ axis.  However, in just the same way as previously discussed, \cite{Nie2014} the disorder leaves us with a nematic transition, represented by the solid black line in the figure, that survives up to a critical value of $\sigma$, defined as the point at which the nematic critical temperature goes to 0.  Strong local CDW correlations persist to finite disorder, vanishing with increasing $\alpha$ or $\sigma$ via a crossover that can be roughly identified with the nematic transition line although surely some local correlations survive arbitrarily far beyond this.  No other symmetries are broken upon decreasing $\alpha$, so what was $T_{CDW2}$ at $\sigma=0$ becomes a crossover line,  (which is only sharply defined for arbitrarily weak disorder) below which a second component of the local CDW correlations should be apparent. (This is indicated by the dashed blue line in the figure.)
 
  Finally, we have indicated two possible Bragg-glass transitions in the figure as solid red lines.  A Bragg glass  is characterized by quasi-long-range (power law) order and an absence of any free dislocations.\cite{Giamarchi1995,gingras1996}  It is thermodynamically distinct from a CDW phase (which has long-range-order) and from a fully disordered phase (which has exponentially falling CDW correlations).  The existence or not of such a phase is not inevitable, as it depends on an appropriate hierarchy of energies (e.g. a large core energy for the dislocation) but it is thought to be possible in 3-dimensions.  In the present case there could be two such phases - one a ``stripe-glass phase'' (discussed previously in Ref.~\cite{Kivelson2000}) in which there is unidirectional CDW quasi-long-range order, and a ``bidirectional glass phase'' which has two orthogonal, non-equivalent CDW correlations, both with quasi-long-range order.  In principle, such quasi-long-range order could be inferred from X-ray diffraction as it leads to a charge-order peak which has a power-law singularity, which is distinct from  the delta-function that rises from long-range-order or the Lorenzian (or squared Lorenzian) that characterizes the disordered phase.  While this distinction is likely difficult to establish in practice, a more promising way to establish such a phase is from STM studies in which disolocations can be directly visualized.  

\end{document}